\documentclass[twocolumn,english,aps,prb,twocolum,superscriptaddress,natbib,bibnotes,amsmath,amssymb,floatfix,groupedaddress,footinbib]{revtex4-2}

\usepackage[english]{babel}
\usepackage{amsmath,amsfonts,amssymb}
\usepackage[T1]{fontenc}
\usepackage{xurl}

\usepackage{amsmath}
\usepackage{siunitx}
\usepackage[version=4]{mhchem}
\usepackage{amsfonts}
\usepackage{amssymb}

\usepackage{epstopdf}
\usepackage{graphicx}
\graphicspath{{./Figures/}}
\usepackage[utf8]{inputenc}

\usepackage[colorlinks=true,citecolor=blue,linkcolor=magenta]{hyperref}

\newcommand{\LN}[0]{$\mathrm{LiNbO}_3$ } 
\newcommand{\LT}[0]{$\mathrm{LiTaO}_3$ } 
\newcommand{\SiO}[0]{$\mathrm{SiO}_2~$ }

\begin{document}
		\title{Arrayed waveguide gratings in lithium tantalate integrated photonics}
	
	\author{Shivaprasad U. Hulyal}
	\affiliation{Institute of Physics, Swiss Federal Institute of Technology Lausanne (EPFL), CH-1015 Lausanne, Switzerland}
	\affiliation{Institute of Electrical and Micro engineering, Swiss Federal Institute of Technology, Lausanne (EPFL), CH-1015 Lausanne, Switzerland}

	\author{Jianqi Hu}
	\affiliation{Institute of Physics, Swiss Federal Institute of Technology Lausanne (EPFL), CH-1015 Lausanne, Switzerland}
	\affiliation{Institute of Electrical and Micro engineering, Swiss Federal Institute of Technology, Lausanne (EPFL), CH-1015 Lausanne, Switzerland}

	\author{Chengli Wang}
	\affiliation{Institute of Physics, Swiss Federal Institute of Technology Lausanne (EPFL), CH-1015 Lausanne, Switzerland}
	\affiliation{Institute of Electrical and Micro engineering, Swiss Federal Institute of Technology, Lausanne (EPFL), CH-1015 Lausanne, Switzerland}
	
	\author{Jiachen Cai}
	\affiliation{Institute of Physics, Swiss Federal Institute of Technology Lausanne (EPFL), CH-1015 Lausanne, Switzerland}
	\affiliation{Institute of Electrical and Micro engineering, Swiss Federal Institute of Technology, Lausanne (EPFL), CH-1015 Lausanne, Switzerland}
	
	\author{Grigory Lihachev}
	\affiliation{Institute of Physics, Swiss Federal Institute of Technology Lausanne (EPFL), CH-1015 Lausanne, Switzerland}
	\affiliation{Institute of Electrical and Micro engineering, Swiss Federal Institute of Technology, Lausanne (EPFL), CH-1015 Lausanne, Switzerland}
	
	\author{Tobias J. Kippenberg}
	\email[]{tobias.kippenberg@epfl.ch}
	\affiliation{Institute of Physics, Swiss Federal Institute of Technology Lausanne (EPFL), CH-1015 Lausanne, Switzerland}
	\affiliation{Institute of Electrical and Micro engineering, Swiss Federal Institute of Technology, Lausanne (EPFL), CH-1015 Lausanne, Switzerland}
	\medskip
	\maketitle

	\noindent\textbf{Arrayed Waveguide Gratings (AWGs) are widely used photonic components for splitting and combining different wavelengths of light. They play a key role in wavelength division multiplexing (WDM) systems by enabling efficient routing of multiple data channels over a single optical fiber \cite{chandrasekhar1995monolithic} and as a building block for various optical signal processing, computing, imaging, and spectroscopic applications \cite{metcalf2016integrated,cheng2024direct,rank2021toward,gatkine2017arrayed}. Recently, there has been growing interest in integrating AWGs in ferroelectric material platforms, as the platform simultaneously provide efficient electro-optic modulation capability and thus hold the promise for fully integrated WDM transmitters. To date, several demonstrations have been made in the X-cut thin-film lithium niobate ($\mathrm{LiNbO}_3$) platform \cite{prost2018compact,yu2022wavelength,wang_chip_2024,tu_100-channel_2024,yi_anisotropy-free_2024,wang2025electro}, 
    yet, the large anisotropy of $\mathrm{LiNbO}_3$ complicates the design and degrades the performance of the AWGs. To address this limitation, we use the recently developed photonic integrated circuits (PICs) based on thin-film lithium tantalate ($\mathrm{LiTaO}_3$) \cite{wang_lithium_2024}, a material with a similar Pockels coefficient as \LN but significantly reduced optical anisotropy, as an alternative viable platform. In this work, we manufacture $\mathrm{LiTaO}_3$ AWGs using deep ultraviolet lithography on a wafer-scale.
    The fabricated AWGs feature a channel spacing of 100 GHz, an insertion loss of <~4~dB and crosstalk of <-14~dB. 
    In addition, we demonstrate a cyclic AWG, as well as a multiplexing and demultiplexing AWG pair for the first time on \LT platform. 
    The wafer-scale fabrication of these AWGs not only ensures uniformity and reproducibility, but also paves the way for realizing volume-manufactured integrated WDM transmitters in ferroelectric photonic integrated platforms.}

	\begin{figure*}[htbp!]
		\centering
		\includegraphics[width=1\linewidth]{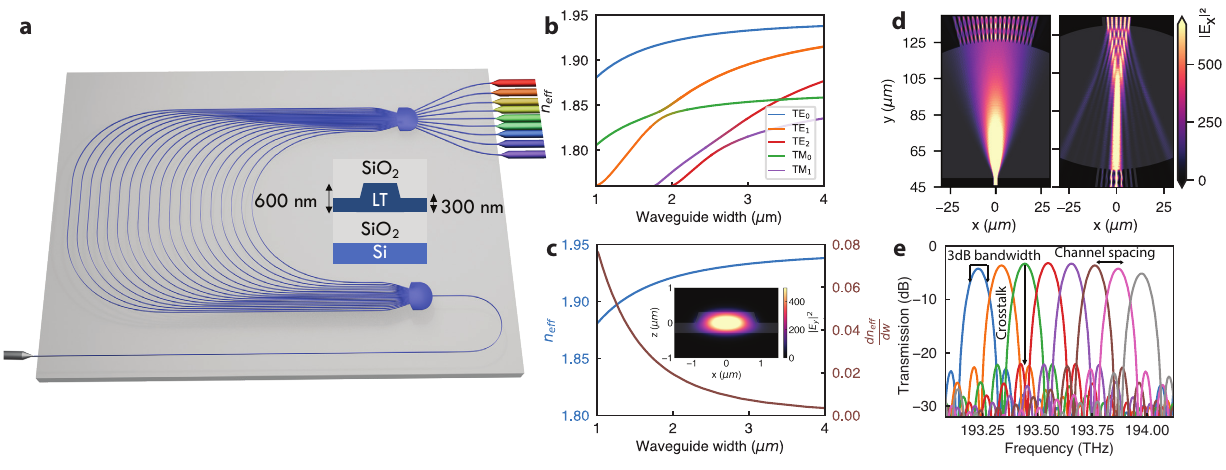}
		\caption{\textbf{Design and simulation of arrayed waveguide gratings based on \LT  photonic integrated circuits.}
			(a) Schematic diagram of the AWG. The inset shows the material stack.
			(b) Refractive index of various modes supported by partially etched \LT waveguide.
			(c) Effective indices and the change in refractive indices at different waveguide widths of the TE$_0$ mode on \LT waveguide.
			(d) (left panel) Simulation of light diffraction from the input waveguide into the star coupler, followed by coupling into arrayed waveguides, (right panel) Simulation of the light entering the output star coupler, after accumulating phase delays in the arrayed waveguides and constructively interfering at one of the output waveguides (the central waveguide is coupled for in-phase illumination).
			(e) Simulated transmission spectrum of the AWG.
			}
		\label{fig1}
	\end{figure*}

	\section{Introduction}

    \noindent Integrated photonics leverages the broad bandwidth of optics, spanning from the visible to the mid-infrared, to enable advanced technologies such as optical telecommunications \cite{hu2018single}, precision metrology \cite{picque2019frequency}, and photonic computing \cite{feldmann2021parallel}. Central to these applications are integrated components like wavelength division multiplexers (WDMs), which efficiently separate, route, and combine multiple optical signals \cite{ishio1984review}.
	For instance, WDMs can be used as both multiplexers and demultiplexers for data transmission of densely parallel channels over optical fibers, thereby significantly enhancing bandwidth utilization and network capacity. Among others, arrayed waveguide gratings (AWGs), have become a building block due to their versatility and efficiency in wavelength routing and spectral analysis \cite{smit_phasar-based_1996, smit1988new}. They have been used for various photonics-based applications, ranging from communications \cite{chandrasekhar1995monolithic}, imaging \cite{rank2021toward}, spectroscopy \cite{gatkine2017arrayed}, sensing \cite{sano2003fast}, to microwave photonics \cite{metcalf2016integrated} and photonic computing \cite{cheng2024direct}. 
    
    Over the past decades, AWGs have been demonstrated on various photonic integrated platforms, including silica \cite{himeno_silica-based_1998}, silicon \cite{yebo_-chip_2011}, silicon nitride \cite{piels_low-loss_2014}, polymer \cite{keil2001athermal}, and indium phosphide \cite{barbarin2004extremely}. More recently, significant efforts have been made to integrate AWGs in the lithium niobate on insulator (LNOI) platform \cite{prost2018compact,yu2022wavelength,wang_chip_2024,tu_100-channel_2024,yi_anisotropy-free_2024,wang2025electro}, leveraging its exceptional electro-optic modulation capability \cite{zhu2021integrated,boes2023lithium}. X-cut lithium niobate ($\mathrm{LiNbO}_3$) is of particular interest, due to its ability to support low driving voltage, high-bandwidth electro-optic modulators \cite{wang_integrated_2018,xu2022dual}. However, in the X-cut $\mathrm{LiNbO}_3$, the optical waveguides experience large orientation-dependent refractive indices.
    Such anisotropy poses challenges in realizing high-performance AWGs in the X-cut. To this end, aligning the symmetry axis of the AWGs along the 45$^{\circ}$ crystal axis has been shown to alleviate the issue by canceling the anisotropy~\cite{yi_anisotropy-free_2024}. Yet, this approach may complicate the overall layout and routing of photonic circuits, potentially increasing the device footprint. Any deviation from the optimal alignment may also reintroduce the anisotropic behavior, requiring tighter fabrication tolerances. 

    Recently, tightly confined photonic integrated circuits with low loss have been demonstrated in lithium tantalate on insulator (LTOI), using a diamond like carbon hardmask (DLC) \cite{wang_lithium_2024}. Compared to $\mathrm{LiNbO}_3$, the integrated lithium tantalate ($\mathrm{LiTaO}_3$) platform offers similar Pockels coefficient, while benefiting from lower wafer per cost due to its large-scale production in consumer electronics \cite{ballandras2019new}.
    High speed optical modulators for communications have been demonstrated in this platform \cite{wang2024ultrabroadband},
    with reduced bias drift \cite{powell2024dc,wang2024ultrabroadband}, higher damage threshold and wider bandgap. 
    Most importantly, \LT features more than ten-fold reduction in optical anisotropy relative to $\mathrm{LiNbO}_3$ ($\Delta n_{\mathrm{LT}}=0.004$ versus $\Delta n_{\mathrm{LN}} = -0.07$). 
    This property of \LT has facilitated the generation of ultra-broadband electro-optic frequency combs beyond the birefringence limit \cite{zhang2025ultrabroadband}, and enables design of compact photonic integrated circuits with tight bend radii. 
    
   Here, by leveraging the reduced anisotropy of $\mathrm{LiTaO}_3$, we demonstrate AWGs on X-cut \LT without having to rely on rotational schemes or other intricate methods to counterbalance anisotropy. We fabricate AWGs for the first time at wafer scale on a ferroelectric material platform, achieving dense channel spacing, competitive insertion loss, and acceptable crosstalk and uniformity across different regions of the wafer. We use both Confocal and Rowland types of star couplers \cite{dragone2002n, munoz_modeling_2002}, which render comparable AWG performance. 
    In addition, we demonstrate a cyclic AWG, as well as a matched multiplexing and demultiplexing AWG pair for the first time to our knowledge. 
    The wafer-scale fabrication of AWGs in a ferroelectric platform provides a scalable and cost-effective solution for on-chip optical communication and spectroscopic systems.

	\section{Design and simulation}
	Figure~\ref{fig1}a illustrates the schematic architecture of an AWG. At its core, an AWG comprises of an input star coupler, an array of waveguides with uniformly incremental path-length differences, and an output star coupler. Initially, the optical signal is injected into the input star coupler, where it undergoes in-plane diffraction, resulting in a fan-out of the light coupling into a number of radial paths. 
    Each path then propagates through an individual waveguide in the array, where a constant length increment is meticulously maintained. This accumulated path-length difference imparts a progressive delay to the different copies of propagating light.
    Upon emerging from the waveguide array, the light enters the output star coupler, where the varying phases accumulated along different waveguides interact to produce constructive and destructive interference patterns. The resulting interference is such that light at specific wavelengths constructively interferes at designated output waveguides, effectively demultiplexing the spectral components of the input signal. 
    Based on the arrangement of the waveguide apertures at the focal plane, the output star coupler can be classified into two primary configurations: Confocal and Rowland-type geometries. In the Confocal geometry, both the input and output waveguides are positioned along a circle with the same radius. While the Rowland geometry places the output waveguides on a circle with half that radius.  In this work, both geometries are designed and fabricated in the thin-film \LT platform.

	\begin{figure*}[htbp!]
		\centering
		\includegraphics[width=1\linewidth]{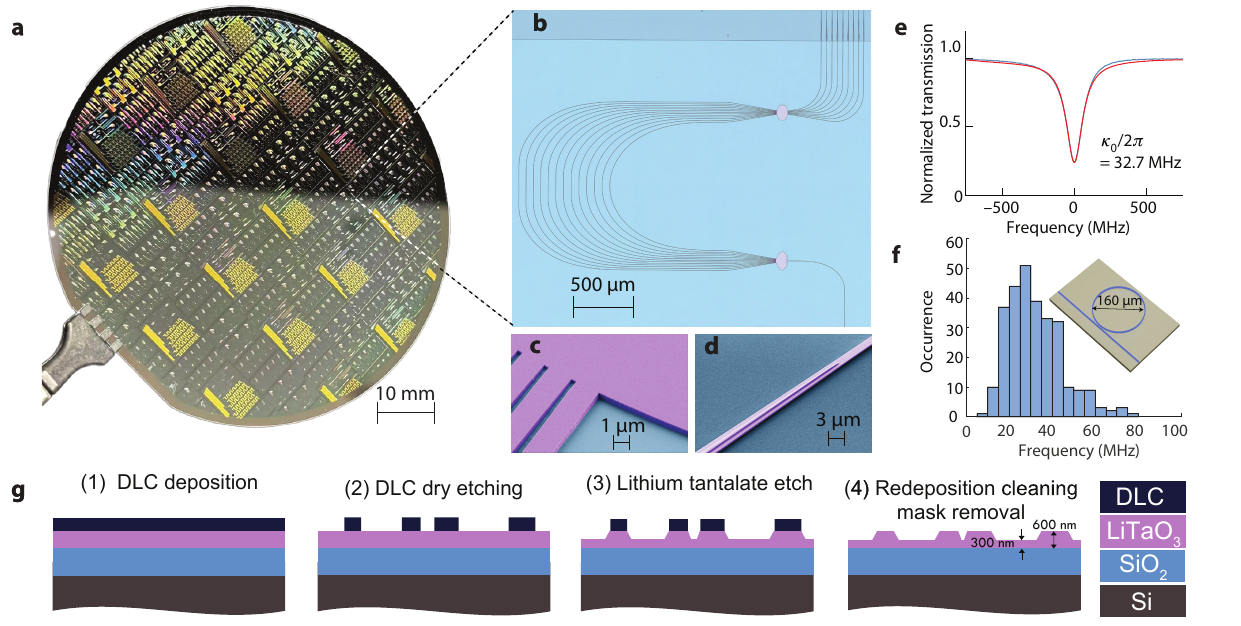}
		\caption{\textbf{Wafer-scale manufacturing of LTOI arrayed waveguide gratings.}
			(a) Optical image of a 4-inch wafer of \LT integrated AWG following the fabrication processes.
			(b) Photo of the photonic chip comprising of the fabricated Confocal type configuration AWG.
			False-colored Scanning Electron Microscope (SEM) images of the arrayed aperture spacing,
			(c) arrayed waveguide aperture, (d) Double layered tapers.
			(e) Normalized resonance transmission spectrum of an optical microring resonator at 204.149 THz (blue curve: measured data; red curve: fitted response).
			(f) Statistical distribution of intrinsic loss, ${\kappa_{0}/2\pi}$. Mean (${\overline{\kappa_{0}}/2\pi}$) = 32.7 MHz, of the optical microring resonator.
			(g) Wafer-scale fabrication process flow. }
		\label{fig2}
	\end{figure*}

	The design parameters of an AWG—including focal length, delay length, free spectral range, dispersion, and diffraction order—are inherently interdependent (cf. Supplementary note 1). These parameters are intricately linked through the effective and group indices of the waveguides, rendering the AWG design process complex. The design of AWG starts by choosing an appropriate waveguide width. A waveguide width whose refractive index is not sensitive to fabrication imperfections nor mode mixing is used. From Figure~\ref{fig1}b we choose a waveguide width in the multimode regime to minimize phase errors caused by sidewall roughness \cite{zhang_reduction_2007}. Figure~\ref{fig1}c shows that the TE$_0$ is tolerant to the waveguide width variations for larger waveguide widths and hence we chose widths of 1.6 $\mathrm{\mu m}$ and 1.8 $\mathrm{\mu m}$ for the Confocal and Rowland AWGs, respectively. 
    Additionally, we designed parabolic taper connecting the waveguides to the star coupler aperture (cf. Supplementary note 3), so as to achieve low-loss mode transition~\cite{ye_low-crosstalk_2014}. These tapers play an essential role in efficiently coupling light between waveguides with different cross-sectional dimensions.  The arrayed waveguides are uniformly bent with identical radii to ensure they remain parallel, and straight sections are subsequently added so that the outermost waveguides extend further than the inner ones.  In our design, Euler bends~\cite{ji_compact_2022} are incorporated into the arrayed waveguides to optimize the optical routing while mitigating propagation losses. Specifically, these bends are engineered with curvature radii of more than 300 $\mathrm{\mu m}$, ensuring a smooth transition that maintains mode integrity and minimizes scattering effects. After that, there are again straight sections to make the appropriate incremental length offset. These waveguides are repeated to reach the output array aperture. In the output star coupler, interference between the optical fields determines the distribution of light into the output waveguides. For a given frequency, the superposition of phase-shifted waves results in constructive interference at specific locations, corresponding to the output channel positions.

	The simulation of AWG (cf. Supplementary note 2) involves both analytical approaches and numerical modeling based on the finite-difference time-domain (FDTD) method \cite{smit_phasar-based_1996, munoz_modeling_2002}. First, the light diffraction from the input star coupler and its coupling into arrayed waveguides are simulated using FDTD in Figure~\ref{fig1}d (left). Next, the light in each arrayed waveguide experiences a precise group delay determined by the calculated length difference between the waveguides. Finally, FDTD is used again to simulate the light propagation in the output star coupler. As shown in Figure~\ref{fig1}d (right), at the center frequency where the light in the arrayed waveguides is in-phase, the light is mainly focused into the central output waveguide with slight leakage into adjacent waveguides leading to crosstalk. Depending on the input light frequency, the light is directed into different waveguides at the output aperture to form the transmission spectrum as shown in Figure~\ref{fig1}e. This figure also illustrates the definitions of 3 dB bandwidth, channel spacing, and crosstalk for the AWG used in this work.

	\begin{figure*}[ht!]
		\centering
		\includegraphics[width=1\linewidth]{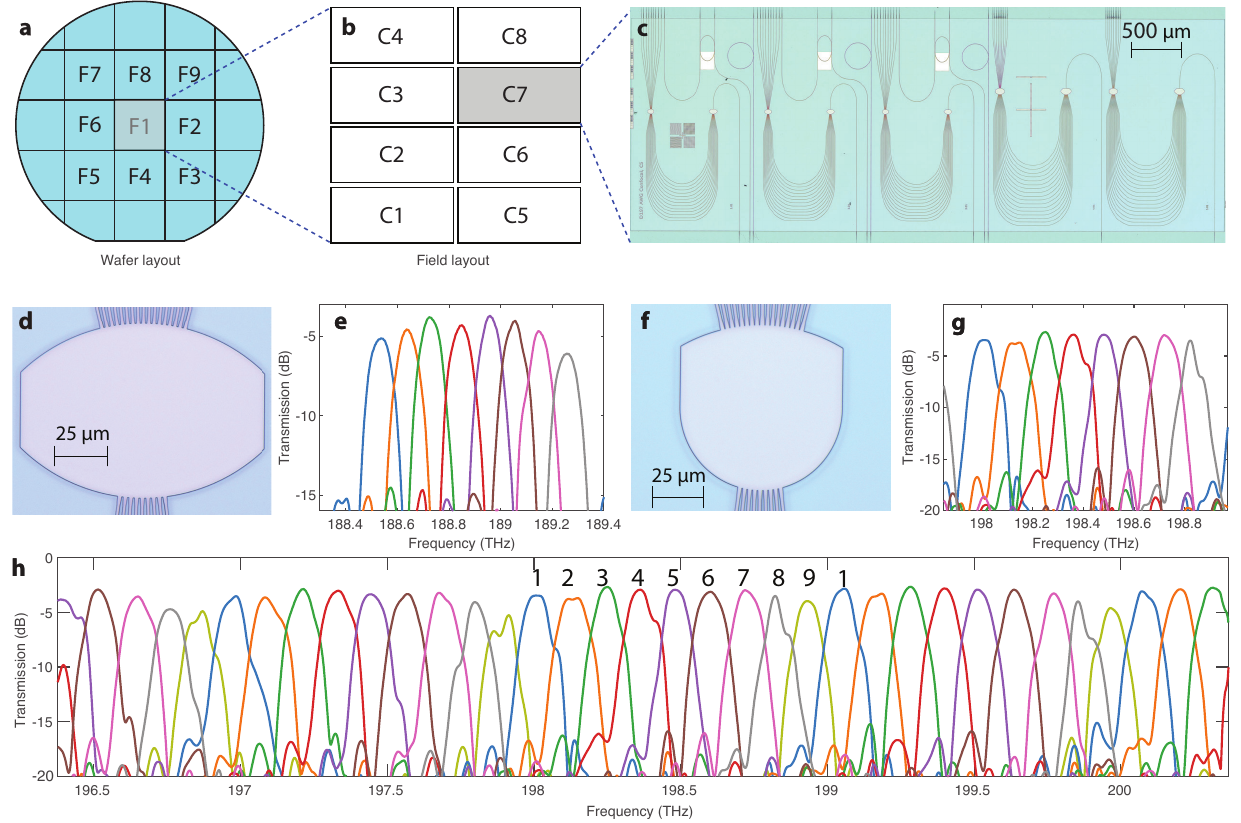}
		\caption{\textbf{Characterization of fabricated \LT arrayed waveguide gratings.}
			(a) DUV stepper exposure layout.
			(b) The reticle design containing eight chips  is uniformly exposed in discrete fields over a 4-inch wafer.
			(c) Photonic image of the fabricated chip containing five confocal 8-channel 100 GHz AWGs in a single chip.
			(d) Microscopic image of fabricated Confocal output star coupler.
			(e) Transmission spectrum of the 8-channel 100 GHz AWG (Device ID: \texttt{D197\textunderscore F5\textunderscore C1\textunderscore WG\textunderscore 203}) with Confocal star couplers.
			(f) Microscopic image of fabricated Rowland output star coupler.
			(g) Transmission spectrum of the 8-channel 100 GHz AWG with Rowland star couplers (Device ID: \texttt{D197\textunderscore F1\textunderscore C6\textunderscore WG\textunderscore 301}).
			(h) Cyclic AWG design with 1 $\times$ 9 x 100 GHz (Device ID: \texttt{D197\textunderscore F1\textunderscore C6\textunderscore WG\textunderscore 301}).
			}
		\label{fig3}
	\end{figure*}

	\section{Fabrication}
	A 4-inch (102 mm) \LT wafer (Figure~\ref{fig2}a) with a surface roughness of 0.25 nm and a non-uniformity of less than 30 nm is used to fabricate the AWGs (Figure~\ref{fig2}b) \cite{wang_lithium_2024}. The LTOI wafer stack comprises a 600 nm single-crystal \LT layer, a 4.7 $\mathrm{\mu m}$ thermal \SiO buried oxide, and a 625 $\mathrm{\mu m}$ high-resistivity ($10~k\Omega\cdot {\rm cm}$) silicon handle substrate. The \LT photonic chips are fabricated based on 248~nm DUV stepper lithography (ASML PAS5500/350C) using a highly selective hardmask subtractive waveguide manufacturing process, with the DLC as the hardmask \cite{wang_lithium_2024} (Figure~\ref{fig2}g). The process begins with the DLC hard mask deposition. A multilayer hard mask stack is deposited via plasma-enhanced chemical vapor deposition (PECVD), consisting of a 30 nm Si\textsubscript{3}N\textsubscript{4} layer, a 480 nm DLC layer, and a top 60 nm Si\textsubscript{3}N\textsubscript{4} layer. Waveguide patterns are initially transferred onto the DLC hard mask using an oxygen plasma etch in a reactive ion etching (RIE) system, taking advantage of the DLC's superior etch selectivity and robustness. Subsequently, the waveguide cores are etched into the \LT layer using ion beam etching (IBE) in a Veeco Nexus IBE350 system. This step achieved an etch depth of approximately 300 nm, retaining a 300 nm-thick slab that is readily compatible with high-speed electro-optic modulation \cite{wang_ultrabroadband_2024}. During IBE, amorphous re-deposited material may form on the sidewalls, and this residue is removed by a post-etch chemical cleaning step to ensure smooth and low-loss waveguide profiles as shown Figure~\ref{fig2}c. An 800 nm \SiO cladding layer is deposited on the optical layer via a ICP-CVD system (PlasmaPro 100 ICPCVD). Device separation is carried out through a multi-step release process: dry etching of the \LT and \SiO layers using fluorine-based chemistry, followed by deep reactive-ion etching of the silicon carrier. This sequence yields smooth end facets suitable for efficient edge coupling via the double layer tapers to external light sources as depicted in Figure~\ref{fig2}d. The double layer tapers are designed to enhance the transmission efficiency (3.7 dB fiber-to-fiber) by matching the mode fields of the lensed fiber and the taper (cf. Supplementary note 5).

	\begin{figure*}[ht!]
		\centering
		\includegraphics[width=\linewidth]{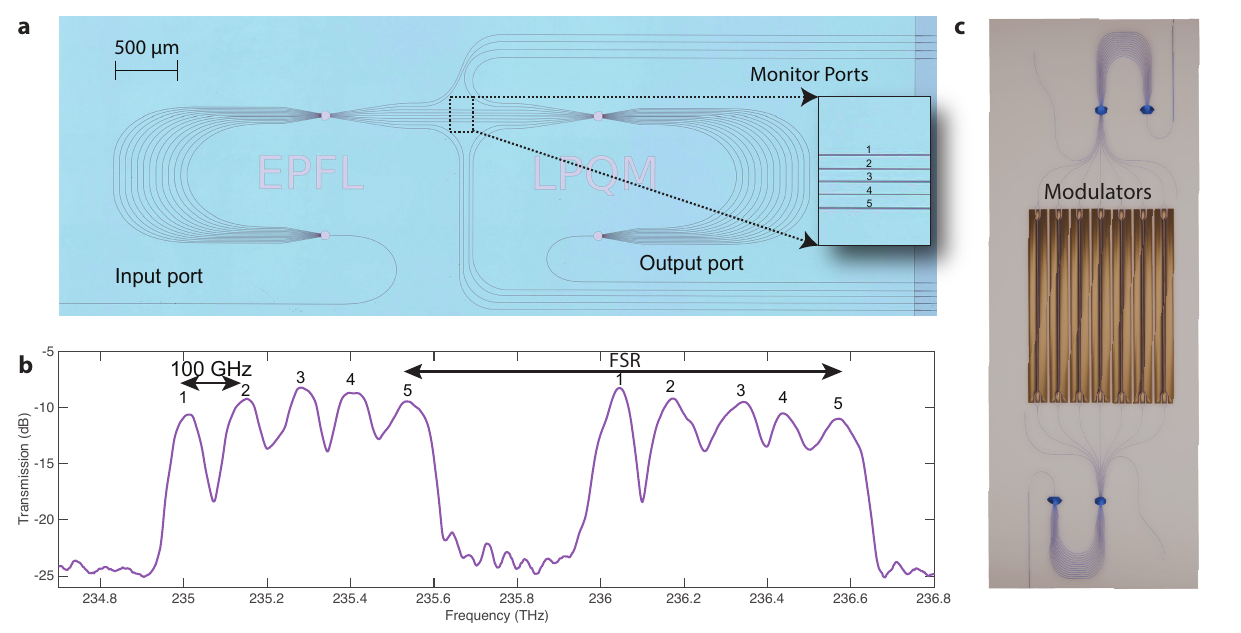}
		\caption{\textbf{Demonstration of multiplexing demultiplexing using arrayed waveguide grating pair.}
		(a) Microscope image of the fabricated (Device ID: \texttt{D197\textunderscore F6\textunderscore C7\textunderscore WG\textunderscore 305}) passive cascaded configuration of two AWGs demonstrating spectral routing. The five outputs of the left AWG are directly connected to the five inputs of AWG on the right, creating a testbed for evaluating passive device characteristics. The inset shows the five ports. 
		(b) Comb-calibrated characterized response showing the clear peaks corresponding to the five channels. 
		(c) Concept of a fully integrated transmitter with modulators between two AWG to encode the data on \LT platform.
   		}
		\label{fig4}
	\end{figure*}

	\section{Characterization}    
    After fabrication, we characterized the \LT AWGs using tunable diode lasers calibrated with a self-referenced frequency comb (Menlo OFC-1500). To ensure a linear frequency scanning, we use the beating between the laser and the comb as a reference to calibrate the laser scan \cite{liu2016frequency}. 
    The propagation loss of \LT waveguides are obtained by measuring the intrinsic loss rate $\kappa_0/2\pi$ of a \LT microresonator in the same wafer, which gives $\overline{\kappa_0}/2\pi$ = 32.7 MHz corresponding to a propagation loss of 5.95 dB/m (Figure~\ref{fig2}e and f).
    Figure~\ref{fig3}a-c show the wafer and chip layout as well as an optical microscope image of the fabricated 8-channel AWGs with Confocal (Figure~\ref{fig3}d) and Rowland (Figure~\ref{fig3}f) configurations, both designed for a nominal 100 GHz channel spacing. The corresponding measured output spectra for the Confocal AWG are presented in Figure~\ref{fig3}e, exhibiting an insertion loss of 6.14 ± 0.72 dB, 3-dB bandwidth of 101.92 ± 7.54 GHz, adjacent channel crosstalk of -13.38 ± 1.64 dB, and a measured channel spacing of 106.73 ± 8.38 GHz, where all uncertainties represent one standard deviation. In comparison, the spectra of the Rowland-type AWG, shown in Figure~\ref{fig3}g, demonstrate an insertion loss of 3.16 dB ± 0.35, 3-dB bandwidth of 92.14 ± 16.86 GHz, adjacent channel crosstalk of -14.67 ± 1.25 dB, and a channel spacing of 116.46 ± 13.58 GHz, also reported with one standard deviation uncertainty. These spectral responses are consistently observed across multiple fields of the wafer (cf. Supplementary note 4), underscoring the excellent reproducibility of AWG performance at the wafer-scale. Furthermore, the measured results are in good agreement with the simulated spectra (Figure~\ref{fig1}e), validating the effectiveness of the design and fabrication methodologies for both configurations.
    The AWG with the Rowland star coupler (Figure~\ref{fig3}g) also exhibits the characteristics of a so-called "cyclic" or "colourless" AWG (Figure~\ref{fig3}h) \cite{seyringerchen_arrayed_2016, dragone_efficient_1989}. That is, a signal that initially exits through the end output channel will reappear at the first output channel when its frequency is increased by an amount equal to the channel spacing of the AWG. Such a cyclic behaviour can be leveraged for a range of applications, particularly for 1 $\times$ 9 wavelength routing and optical add-drop multiplexers.

	Next, we implement a pair of identical AWGs interconnected as shown in Fig.~\ref{fig4}a. Specifically, five channels from the first AWG are routed to the second AWG, while remaining channels serve as monitoring ports. Figure~\ref{fig4}b shows the characterization results of the device, clearly showing the spectral response of the five channels. Precise fabrication is crucial for this configuration to function properly; any deviations can lead to misalignment in the AWG transmission response, resulting in no detectable output. This multiplexing-demultiplexing configuration is a key step towards realizing a fully integrated \LT WDM transmitter,
    as illustrated in Figure~\ref{fig4}c, providing a path for future Tb/s-scale optical communication system.

	\section*{Conclusion}
    In this work, we report the first demonstration of AWGs on thin-film \LT based photonic integrated circuit platform.     
    Notably, these AWGs operate passively, requiring neither active control nor complex design, but solely relying on the inherently low anisotropy of the material. Fabricated on X-cut \LT that supports high-speed modulators, the AWGs achieve a channel spacing around 100 GHz, which aligns with the ITU grid standard and thereby paves the way for highly dense WDM transmitters on chip. Toward this goal, wafer-scale fabrication of AWGs is desired, offering a scalable, reproducible and cost-effective solution for volume manufacturing. 

    Besides its use in optical communications, the AWGs in X-cut LTOI platform also hold potential for many other photonic-based applications. Since superior electro-optic frequency combs have been developed in the ferroelectric platform \cite{zhang2019broadband,yu2022integrated,zhang2025ultrabroadband}, AWGs can combine monolithically with the comb, empowering on-chip comb-based microwave and optical signal processing \cite{metcalf2016integrated}, light ranging and detection \cite{riemensberger2020massively}, photonic computing \cite{feldmann2021parallel}, and quantum information processing \cite{lukens2016frequency}. In addition to the resonant electro-optic combs \cite{zhang2025ultrabroadband} and AWGs demonstrated in this work, the reduced anisotropy of \LT may also facilitate the development of other linear and nonlinear functionalities in this platform, e.g., the dispersion management module \cite{liu2024ultra} and stimulated Brillouin scattering \cite{ye2024brillouin}. These building blocks would propel photonic systems in LTOI to new heights for next-generation communication and computing technologies.

	\begin{footnotesize}
		
		\noindent \textbf{Note:}  During the preparation of this manuscript another competing paper \cite{huang_toward_2025} was published.

		\noindent \textbf{Author Contributions}:
		S.U.H. simulated and designed the devices.
		C.W., J.C. fabricated the device.
		S.U.H. carried out the measurements and analyzed the data with the help of J.H. 
		T.J.K., J.H. supervised the project.
		
		\noindent \textbf{Funding Information and Disclaimer}: This work was supported by Contract N660012424006 (NaPSAC) from the Defense Advanced Research Projects Agency (DARPA), Defense Sciences Office (DSO).  It was further supported by the Swiss National Science Foundation under grant agreement No. 21649 (HEROIC).  S.U.H acknowledges support from Horizon Europe research and innovation programme under the Marie Skłodowska-Curie grant agreement No 101119968 (MicrocombSys) and from the Swiss State Secretariat for Education, Research and Innovation (SERI).

		\noindent \textbf{Acknowledgments}:
		The samples were fabricated in the EPFL Center of MicroNanoTechnology (CMi).
		
		\noindent \textbf{Data Availability Statement}: The code and data used to produce the plots within this work will be released on the repository \texttt{Zenodo} upon publication of this preprint.
		
		\noindent\textbf{Correspondence and requests for materials} should be addressed to T.J.K.
	\end{footnotesize}
	\bibliography{main}
	
\end{document}


\title{Supplementary Information for: Arrayed waveguide gratings in lithium tantalate integrated photonics}

\author{Shivaprasad U. Hulyal$^{1}$,
		Jianqi Hu$^{1}$,
		Chengli Wang$^{1}$,
		Jiachen Cai$^{1}$,
		Grigory Lihachev$^{1}$,
		Tobias J. Kippenberg$^{1,2}$}
\affiliation{
$^1$Institute of Physics, Swiss Federal Institute of Technology Lausanne (EPFL), CH-1015 Lausanne, Switzerland\\
$^2$Institute of Electrical and Micro engineering, Swiss Federal Institute of Technology, Lausanne (EPFL), CH-1015 Lausanne, Switzerland
}

\setcounter{equation}{0}
\setcounter{figure}{0}
\setcounter{table}{0}

\setcounter{subsection}{0}
\setcounter{section}{0}
\setcounter{secnumdepth}{3}

\maketitle
{\hypersetup{linkcolor=blue}\tableofcontents}
\newpage

\section{Analytic simulation of AWG}
\begin{figure*}[htbp!]
	\centering
	\includegraphics[width=1\linewidth]{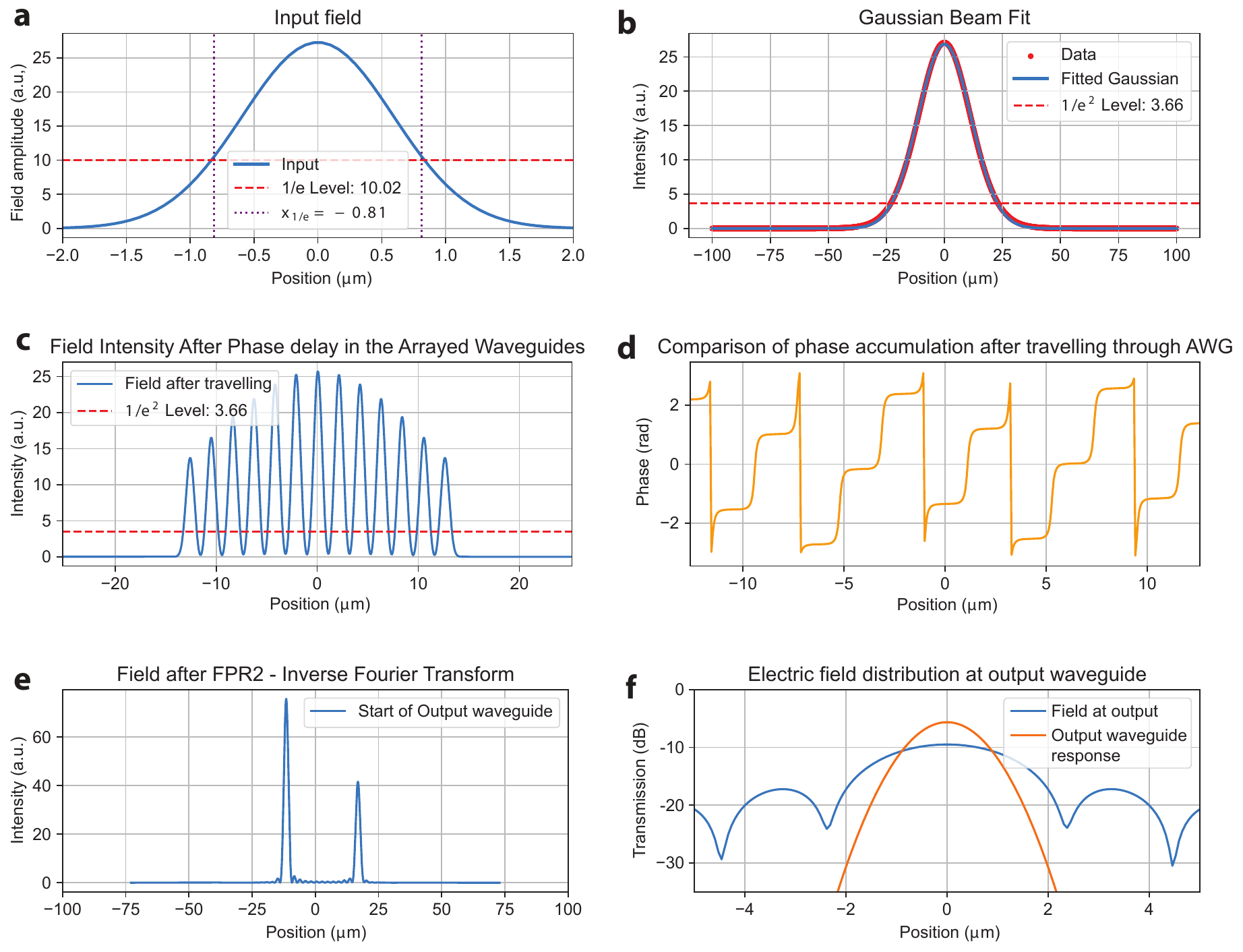}
	\caption{\textbf{Analytical simulation of AWG for the center channel.}
		(a) Gaussian beam at the start of input waveguide. 
		(b) Expanded Gaussian beam after diffraction through the input star coupler. 
		(c) Field intensity after coupling into arrayed waveguide and its corresponding intensity. 
		(d) Phase accumulation after traveling through the delay lines. 
		(e) Field amplitude at the output aperture.
		(f) Overlap between the output field intensity and the waveguide response for the central output waveguide.
		}
	\label{fig1}
\end{figure*}

We use a mathematical model implemented in Python to calculate and analyse the transmission spectrum of AWG.

\subsection{Input Field as a Gaussian Beam}

The input waveguide of an AWG supports a fundamental mode that can be approximated by a Gaussian beam profile as in Figure \ref{fig1}a. This Gaussian field represents the electric field distribution at the aperture of the input star coupler (free propagation region). In a planar waveguide geometry, assuming the mode is symmetric in the transverse $(x,y)$ plane, the field at the star-coupler entrance (position $z=0$) can be written as: 

\begin{equation} 
	E_{\mathrm{in}}(x) = E_\mathrm{0} \exp\Big(-\frac{x^2}{w_\mathrm{0}^2}\Big)
	\label{eq:gaussian} 
\end{equation} 

where $E_\mathrm{0}$ is the peak field amplitude and $w_\mathrm{0}$ is the mode field radius (beam waist). Equation (\ref{eq:gaussian}) is the standard Gaussian beam equation, capturing the fact that the field amplitude has a Gaussian transverse profile.

\subsection{Propagation via Fresnel Diffraction}

As the Gaussian beam propagates a distance \( z_\mathrm{1} \) through the input free propagation region (star coupler), it diffracts and broadens as shown in Figure \ref{fig1}b. The resulting field at the plane of the arrayed waveguides is calculated using the Fresnel diffraction integral:

\begin{equation}
	E(x,z) = \frac{e^{ikz}}{i \lambda z} \int_{-\infty}^{\infty} E_{\mathrm{in}}(x', 0) 
	\exp\left[ \frac{ik (x - x')^2}{2z} \right] dx'
	\label{eq:fresnel}
	\end{equation}
where $E_{\mathrm{in}}(x',0)$ is the field at the input plane (given by the Gaussian of Eq. (\ref{eq:gaussian})), $k = 2\pi/\lambda$ is the wavenumber, and $\lambda$ is the effective wavelength. Equation (\ref{eq:fresnel}) is the Fresnel diffraction formula in one transverse dimension (here $x$) for the field at the array plane located a distance $z$ from the input. This integral accounts for the diffraction spreading of the beam: each point $x'$ in the input aperture contributes a spherical wave that interferes at observation point $x$.
In the paraxial regime, the transformation from the input field to the focal plane field can be modeled as a spatial Fourier transform of the input distribution. \cite{parker_design_1999}
\begin{equation}
	E(x,z) = 1/\sqrt{b} \int_{-\infty}^{\infty} E_{\mathrm{in}}(x', 0)  \exp \left(j 2 \pi x' b x \right) d x'
	\label{eq:fouriertransform}
\end{equation}
where $b$ is the scaling factor given by,
\begin{equation}
	b = \frac{n_\mathrm{s} v_\mathrm{0}}{c z}
	\label{eq:fouriertransform}
\end{equation}

which holds in the region of Fraunhofer approximation.
\begin{equation}
	z \gg \frac{\pi w_\mathrm{0}^2}{4 \lambda}
	\label{eq:largefieldlimit}
\end{equation}

\subsection{Coupling into Arrayed Waveguides}

The diffracted field reaches the apertures of the arrayed waveguides, which are located at discrete positions \( x_n \), where \( n = 1, 2, \dots, N \). The amplitude coupled into the \( n \)-th waveguide is given by the overlap between the diffracted field and the mode profile \( \psi_{\text{wg}}(x) \) of the waveguides:

\begin{equation}
	a_\mathrm{n}(\lambda) = \int_{-\infty}^{\infty} E(x,R_s) \, \psi^*_{\text{wg}}(x - x_\mathrm{n}) \, dx
	\label{eq:coupling}
	\end{equation}
where $R_s$ is the radius of the grating circle.
This integral represents the excitation strength of each waveguide in the array, forming the input to the delay line network and the final field front is plot in Figure \ref{fig1}c.

\subsection{Phase Accumulation in Arrayed Waveguides}
A phase delay term is added to model the accumulated phase after traveling through the set of $N$ delay lines. The pahse accumulated is plotted in Figure \ref{fig1}d.
\begin{equation}
	\Phi_n(\lambda) = \frac{2\pi n_{\text{eff}}}{\lambda} \left( L_0 + (N - 1) \Delta L \right)
	\label{eq:phase}
\end{equation}

where \( n_{\text{eff}} \) is the effective refractive index of the waveguide and  $L_0$ is the initial offset in the total length.

\subsection{Interference in the Output Star Coupler}
Each output waveguide is positioned at a specific $x = X_j$ to capture a particular wavelength's focal spot (channel $j$ corresponds to a central wavelength $\lambda_j$). The coupling into an output waveguide is again given by an overlap integral between the field in the star coupler output plane and the mode of the output waveguide. The beams from the arrayed waveguide interfere with each other in the output coupler and form a combined field distribution at the plane of the output waveguide array.
This is shown in Figure \ref{fig1}e.  Because each waveguide $n$ has an amplitude $a_n$ (from Eq. (\ref{eq:coupling})) and a phase $\Phi_n(\lambda)$ (from Eq. (\ref{eq:phase})), the total electric field in the output coupler is the superposition of all $N$ contributions. We can write the field at a position $x$ in the output focal plane as: 

\begin{equation}
	E_{\text{out}}(x,\lambda) = \sum_{n=1}^{N} a_n(\lambda) \, e^{i \Phi_n(\lambda)}  \psi_{\text{wg}}(x - x_n')
	\label{eq:interference}
	\end{equation}
The output star coupler can be considered to be as the interference from the different phase accumulated in the arrayed waveguides to focus the desired wavelength into the corresponding output waveguide. 

The focused interference pattern in the output plane is collected by output waveguides located at positions \( X_j \), corresponding to wavelength channels. 

\begin{equation}
	c_j(\lambda) = \int_{-\infty}^{\infty} E_{\text{out}}(x,\lambda) \, \psi^*_{\text{wg}}(x - X_j) \, dx
	\label{eq:overlap}
\end{equation}
Figure \ref{fig1}f shows the overlap of the field intensity of output star coupler and the response of a single waveguide.
The amount of power coupled into the \( j \)-th output waveguide is given by the overlap integral:
\begin{equation}
	P_j(\lambda) = |c_j(\lambda)|^2
	\label{eq:power}
\end{equation}

\section{Device parameters selection}
The radius ($R_a$) of the input grating circle is chosen to accommodate the total number of arrayed waveguides which can be hence calculated as,
\begin{equation}
	R_a=\frac{d_a N}{\theta_{d a}}
\end{equation}
where $d_a$ is the spacing between the arrayed waveguides and $\theta_{d a}$ is the width of the Gaussian far field. A larger number of arrayed waveguides reduces the bandwidth of the individual channels whereas a larger gap between the adjacent arrayed waveguides increases the insertion loss.

Efficient focusing of the propagating optical fields is achieved when the path length difference $\Delta L$ between adjacent waveguides corresponds to an integer multiple $m$ of the effective wavelength within the waveguide medium.
\begin{equation}
	\Delta L=m \cdot \frac{\lambda_c}{n_{\mathrm{g}}}
\end{equation}
Here $\Delta L$ is the length difference between adjacent arrayed waveguides and $\lambda_c$ is the center channel wavelength, $n_{\mathrm{g}}$ is the group index of the material at a particular frequency ($v$), given by,
\begin{equation}
	n_{\mathrm{g}}=n_{\mathrm{eff}}+v \frac{\mathrm{~d} n_{\mathrm{eff}}}{\mathrm{~d} v}
\end{equation}
The free spectral range (FSR) is related to the so-called order $m$ of the AWG by and is chosen to fit all the channels of an AWG,
\begin{equation}
	m = \frac{n_{\mathrm{eff}}}{n_{\mathrm{g}}} \frac{v_c}{v_\mathrm{FSR}}
\end{equation}

The dispersion ($D$) of the array defined as the lateral displacement ($\mathrm{d} s$) of the focal spot along the image plane per unit frequency change \cite{smit_phasar-based_1996}, determines the spacing of the output waveguides ($d_o$), 
\begin{equation}
	D =\frac{\mathrm{d} s}{\mathrm{~d} v}=\frac{1}{v_{\mathrm{c}}} \frac{n_{\mathrm{g}}}{n_{\mathrm{FPR}}} \cdot \frac{\Delta L}{\Delta \alpha}
\end{equation}
where $\Delta \alpha$ is the divergence angle of the set of arrayed waveguides, defined as,
\begin{equation}
	\Delta \alpha=d_a / R_a
\end{equation}

The parameters used for the AWG fabricated in the main text are listed in Table \ref{tab:awg_design},
\begin{table}[h]
	\centering
	\begin{tabular*}{\textwidth}{@{\extracolsep{\fill}} lcccccccc}
	\toprule
	\textbf{Configuration} & \boldmath$m$ & \boldmath$N$ & \boldmath$\Delta L$ ($\mu m$) & \boldmath$w_a$ ($\mu m$) & \boldmath$d_a$ ($\mu m$) & \boldmath$R_{\mathrm{a}}$ ($\mu m$) & \boldmath$w_o$ ($\mu m$) & \boldmath$d_o$ ($\mu m$) \\
	\midrule
	Confocal: 1 $\times$ 8 $\times$ 100 GHz & 160 & 13 & 129.7  & 1.6 & 2.1 & 75 & 2.0 & 2.9 \\
	
	Rowland: 1 $\times$ 8 $\times$ 100 GHz & 160 & 13 & 129.0  & 1.8 & 2.3 & 75 & 1.8 & 2.6 \\
	\bottomrule
	\end{tabular*}
	\caption{Design parameters of the fabricated AWG}
	\label{tab:awg_design}
\end{table}

\section{Parabolic taper design}
Within the AWG framework, the implementation of adiabatic tapers is an effective strategy to minimize coupling losses between single-mode waveguides and the free propagation region (FPR), thereby enhancing overall device performance. Here we employ a parabolic taper design with to convert the mode from 2 $\mu m$ to the fundamental mode of 1.6 $\mu m$ (1.8 $\mu m$) for the Confocal (Rowland) type configuration on a 300 nm partially \LT platform over a length of 5 $\mu m$. The parabolic design follows the equation:
	\begin{equation}
		y=a(L_{taper}-x)^n+b
	\end{equation}
where $L_{taper}$ is the length of the taper, and $n$ is the power of the exponent. $a$ and $b$ are two constants determined by input and output waveguide width. $n$ = 2 for the parabolic taper demonstrated an excellent reduced coupling loss at the interface of the star couplers and arrayed waveguides and hence the total insertion loss.
Extensive research in taper design has led to various configurations optimized for efficient mode transformation across multiple photonic platforms, such as silicon \cite{ye_low-crosstalk_2014} and silicon nitride \cite{zhang_ultracompact_2020}.

\section{Wafer scale characterization of AWG}
\begin{figure*}[htbp!]
	\centering
	\includegraphics[width=1\linewidth]{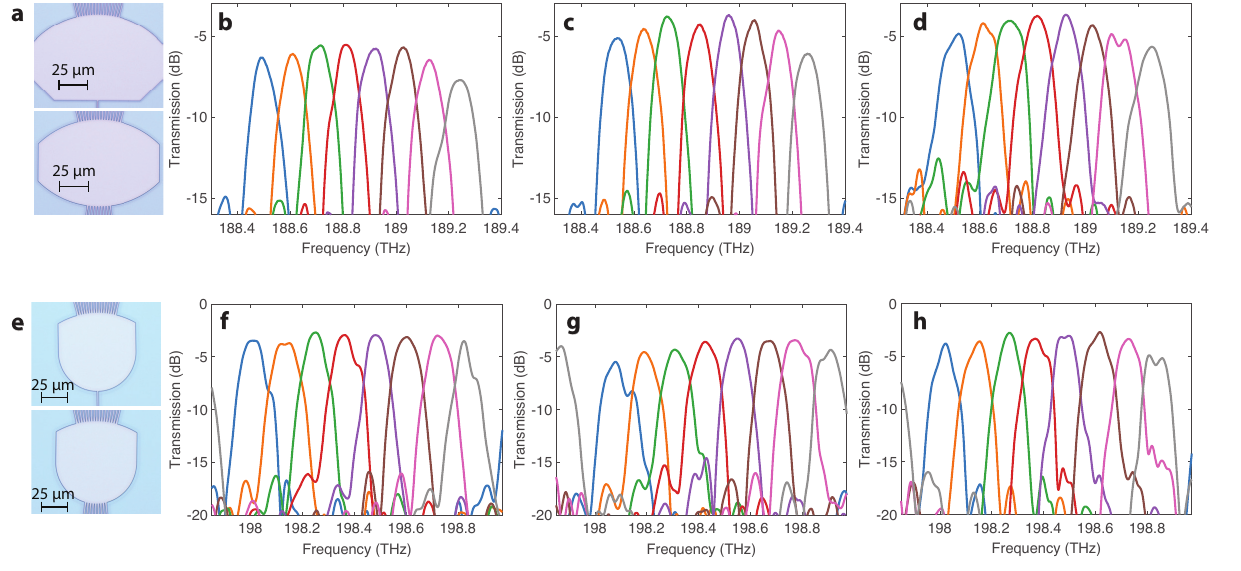}
	\caption{\textbf{Wafer scale AWG characterization results.}
	(a) Microscopic image of fabricated Confocal input star coupler (top) and output star coupler (bottom).
	Transmission spectrum of AWG with Confocal star coupler (Device ID: \texttt{D197\textunderscore C1\textunderscore WG\textunderscore 203}) across different fields of the wafer -
	(b) \texttt{F5},
	(c) \texttt{F1},
	(d) \texttt{F3}.
	(e) Microscopic image of fabricated Rowland input star coupler (top) and output star coupler (bottom).
	Transmission spectrum of AWG with Rowland star coupler (Device ID: \texttt{D197\textunderscore C6\textunderscore WG\textunderscore 301}) across different fields of the wafer -
	(f) \texttt{F1},
	(g) \texttt{F4},
	(h) \texttt{F6}.
	}
	\label{fig2}
\end{figure*}

We characterized the AWGs across multiple chips and fields on the wafer, with the results presented in Figure~\ref{fig2}. For the Confocal AWGs, we measured an average insertion loss of 5.02 ± 1.06 dB, 3-dB bandwidth of 103.75 ± 9.80 GHz, adjacent channel crosstalk of -12.98 ± 3.39 dB, and a measured channel spacing of 104.64 ± 10.59 GHz. All values are reported with one standard deviation. In comparison, the spectra of the Rowland-type AWGs exhibited an insertion loss of 3.53 dB ± 0.69, 3-dB bandwidth of 93.96 ± 15.51 GHz, adjacent channel crosstalk of -14.27 ± 2.13 dB, and a channel spacing of 114.98 ± 14.87 GHz, also reported with one standard deviation uncertainty.

\begin{figure*}[htbp!]
	\centering
	\includegraphics[width=0.5\linewidth]{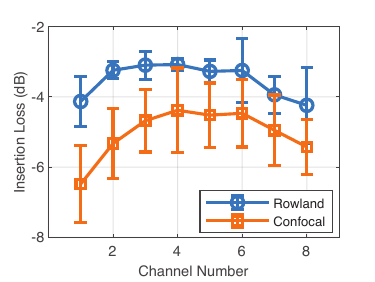}
	\caption{\textbf{Non-uniformity loss of Rowland and Confocal AWGs.}
	Insertion loss plotted as a function of channel number. Error bars represent the standard deviation calculated from measurements across the three AWGs, shown in Figure \ref{fig2}.
		}
	\label{fig3}
\end{figure*}

The non-uniformity loss in AWGs arises from variations in output power across different channels and also from fiber-to-chip coupling. Comparing the two different designs fabricated, Rowland AWGs offer better performance in terms of uniformity (Figure \ref{fig3}).

\section{Double Layer Taper design for AWG}
Efficient optical coupling between the \LT{} waveguide and the mode from the lensed fiber is achieved using double-layered waveguide tapers. In this design, the waveguide width is linearly reduced from the standard width to 0.16~$\mu$m over a length of 160~$\mu$m. Beyond this point, the waveguide is cut off, and the underlying slab continues with a linear taper from its initial width down to 0.16~$\mu$m over an additional 100~$\mu$m, terminating at the chip facet. This taper was patterned using electron beam lithography. This double-tapered structure facilitates low-loss mode matching with the lensed fiber.
The coupling efficiency from the lensed fiber mode and \LT taper mode can be estimated by mode overlap integral:
\begin{equation}
    \eta = \frac{\left|\int_A E^*_\mathrm{lensed \, fiber}\cdot E_\mathrm{taper}dA\right|^2}{\int_A{\left|E_\mathrm{lensed \, fiber}\right|^2dA\cdot \int_A\left|E_\mathrm{taper}\right|^2dA}}, 
\end{equation}
where $E_\mathrm{\mathrm{lensed \, fiber}}$ is the mode field distribution at the lensed fiber  and $E_\mathrm{taper}$ is the field distribution of the \LT taper \cite{puckett2021broadband}.

\begin{figure*}[htbp!]
	\centering
	\includegraphics[width=0.5\linewidth]{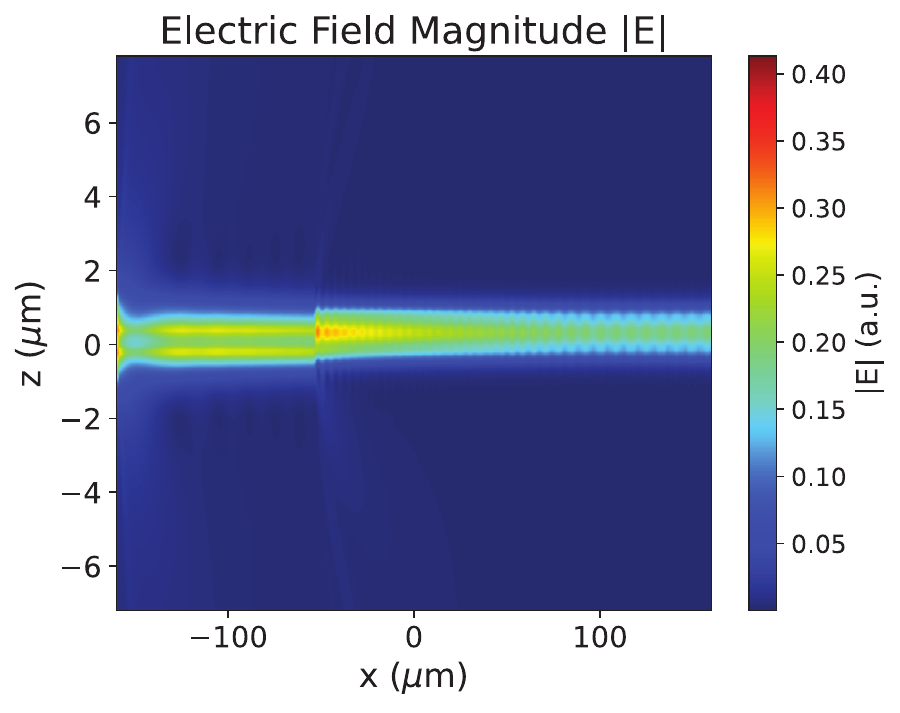}
	\caption{\textbf{FDTD simulations of double layer taper.}
		Electric field distribution showing the coupling of the mode from lensed fiber to the slab and then to the waveguide.
		}
	\label{fig4}
\end{figure*}

We designed double layer tapered edge couplers to couple light into the chip (Figure \ref{fig4}). The taper exhibits a linear transition from the 300 nm slab to the 600 nm waveguide core height over a length of 250 $\rm {\mu m}$.
Edge couplers enable lower insertion losses due to their direct butt-coupling configuration, which minimizes scattering and diffraction losses that are often encountered in grating coupler structures. This direct interface facilitates a more efficient transfer of optical power between the lensed fiber and the edge coupler. 

\bibliographystyle{apsrev4-2}
\bibliography{supp}